\documentclass{ws-p8-50x6-00}

\renewcommand{\rightnote}{{\it March 26, 2002}}

\newcommand{\FF}{{\cal F}_{\pi^0\gamma^*\gamma^*}}
\newcommand{\lag}{{\cal L}}
\newcommand{\order}{{\cal O}}
\newcommand{\lapprox}{%
\mathrel{%
\setbox0=\hbox{$<$}
\raise0.6ex\copy0\kern-\wd0
\lower0.65ex\hbox{$\sim$}
}}

\begin{document}

\def\thefootnote{\fnsymbol{footnote}} 

\title{
\hfill {\normalfont CPT-2002/P.4341} \\
\hfill {\normalfont M\lowercase{arch} 2002} \\[1cm]
Hadronic light-by-light scattering contribution to the muon
$\lowercase{g}-2$ in large-$N_C$ QCD\footnote{T\lowercase{alk 
presented at the} W\lowercase{orkshop on the} P\lowercase{henomenology
of} L\lowercase{arge}-$N_C$ QCD, T\lowercase{empe,}
A\lowercase{rizona}, USA, 9--11 J\lowercase{anuary}
2002. T\lowercase{o be published in the} P\lowercase{roceedings.}}} 

\author{Andreas Nyffeler}

\address{Centre de Physique Th\'{e}orique, CNRS-Luminy, Case 907 \\ 
    F-13288 Marseille Cedex~9, France \\ 
    E-mail: nyffeler@cpt.univ-mrs.fr}


\maketitle

\abstracts{
We present our recent semi-analytical evaluation of the pion-pole
contribution to hadronic light-by-light scattering, using a
description of the pion-photon-photon form factor based on large-$N_C$
and short-distance properties of QCD. Inclusion of all light
pseudoscalar states leads to $a_{\mu}^{\mbox{\tiny{LbyL;PS}}} = +
8.3~(1.2) \times 10^{-10}$. We also sketch an effective field theory
approach to hadronic light-by-light scattering. It yields the leading
logarithmic terms that are enhanced by a factor $N_C$ and shows that
the sign of $a_{\mu}^{\mbox{\tiny{LbyL;had}}}$ cannot be inferred from
a constituent quark-loop. In view of several problematic issues that
still remain to be clarified, our estimate for the full hadronic
light-by-light scattering contribution is
$a_{\mu}^{\mbox{\tiny{LbyL;had}}} = + 8~(4)
\times 10^{-10}$.}

\renewcommand{\thefootnote}{\arabic{footnote}}
\setcounter{footnote}{0}

% --------------------------------------------------------------------

\section{Introduction} 

In early 2001, the Muon $g-2$ Collaboration at the Brookhaven National
Laboratory announced~\cite{BNL_2001} a deviation of 2.6~$\sigma$
between their measured value for $a_\mu$ from the prediction in the
Standard Model (SM).

At the same time, we were working on a description of the
pion-photon-photon transition form factor $\FF$ based on large-$N_C$
QCD, which incorporates short-distance constraints from the operator
product expansion (OPE)~\cite{KN_VAP}. This form factor enters in the
numerically dominant pion-pole contribution to the hadronic
light-by-light scattering correction to $g-2$, denoted by
$a_{\mu}^{\mbox{\tiny{LbyL;had}}}$.~\cite{HKS,BPP} Since the form
factors used in~\cite{HKS,BPP} did not fulfill all short-distance
constraints, we found it worthwhile to reevaluate this pion-pole
contribution $a_{\mu}^{\mbox{\tiny{LbyL;$\pi^0$}}}$.  Furthermore, we
wanted to push the analytical calculation further than was done in
Refs.~\cite{HKS,BPP}, which relied purely on numerical methods.

Our semi-analytical calculation~\cite{KN_pion} produced a result with
roughly the same absolute size but with an opposite sign compared to
the evaluations by three other groups~\cite{HKS,BPP,Bartos_etal}.  In
an accompanying paper~\cite{a_mu_EFT} we therefore provided a
derivation of the coefficient of the leading log-square term, based on
an effective field theory (EFT) approach and the renormalization group
(RG), which lead again to a positive coefficient. The correctness of
the positive result for $a_{\mu}^{\mbox{\tiny{LbyL;$\pi^0$}}}$ was
later confirmed by the authors of Refs.~\cite{HKS,BPP} and by
independent calculations, see Ref.~\cite{confirmation}. The new result
for $a_{\mu}^{\mbox{\tiny{LbyL;had}}}$ reduces the discrepancy between
the experimental value and the SM prediction to about 1.6~$\sigma$.

Below I will present our evaluation~\cite{KN_pion} of
$a_{\mu}^{\mbox{\tiny{LbyL;$\pi^0$}}}$ (Sec.~\ref{sec:pionpole}) and
sketch an EFT approach~\cite{a_mu_EFT} to
$a_{\mu}^{\mbox{\tiny{LbyL;had}}}$
(Sec.~\ref{sec:EFT}). Section~\ref{sec:comments_conclusions} contains
some considerations about leading and next-to-leading logarithms and
an argument why the sign of $a_{\mu}^{\mbox{\tiny{LbyL;had}}}$ cannot
be inferred from a constituent quark-loop. I will conclude with an
estimate for $a_{\mu}^{\mbox{\tiny{LbyL;had}}}$ with a conservative
error bound which takes into account the still unsolved problems.

% --------------------------------------------------------------------

\section{Pion-pole contribution}
\label{sec:pionpole} 

\begin{figure}[h!]
\epsfxsize=18pc % will enlarge or reduce the postscript figures based
%on the xsize 
\centerline{\epsfbox{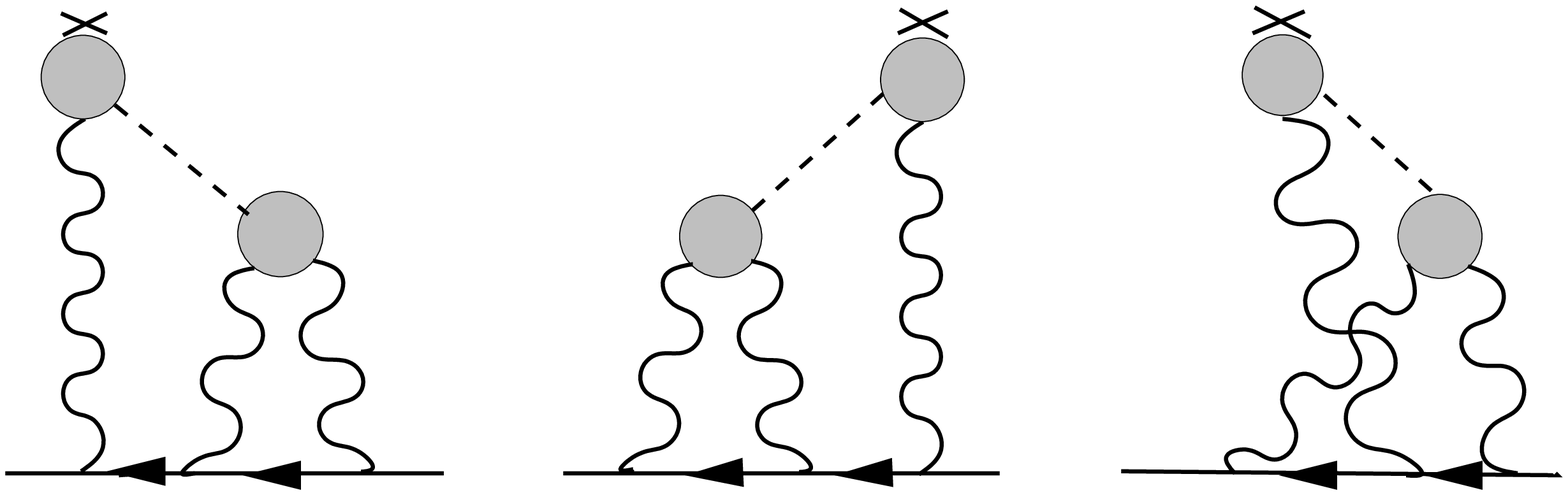}} % postscript image file name
\caption{The pion-pole contributions to light-by-light scattering. The
shaded blobs represent the form factor $\FF(q_1^2,q_2^2)$, defined in
Eq.~(\ref{def_FF}). \label{fig:pionpole}}
\end{figure}

The quantity $a_{\mu}^{\mbox{\tiny{LbyL;had}}}$ involves the
fourth-rank vacuum polarization tensor $\Pi_{\mu\nu\lambda\rho}
(q_1,q_2,q_3)$, built from the light-quark electromagnetic current
$j_{\mu}=(2{\bar u}\gamma_{\mu}u - {\bar d}\gamma_{\mu}d - {\bar
s}\gamma_{\mu}s)/3$. In this section we will concentrate on the
contribution from the neutral pion intermediate state, see
Fig.~\ref{fig:pionpole},
which can be represented as follows~\cite{KN_pion} 
\bea
a_{\mu}^{\mbox{\tiny{LbyL;$\pi^0$}}}& = & - 
e^6\!\!\int\!\!{d^4 q_1 \over (2\pi)^4}\!\!\!\int\!\!{d^4 q_2 \over
(2\pi)^4} \,\frac{1}{q_1^2 q_2^2 (q_1 + q_2)^2[(p+ q_1)^2 - m^2][(p -
q_2)^2 - m^2]} 
\nonumber \\
&& \quad \quad \times \left[ {\FF(q_1^2, (q_1 + q_2)^2) \ \FF( q_2^2,
0) \over q_2^2 - M_{\pi}^2} \ T_1(q_1,q_2;p) \nonumber \right. \\ &&
\quad \quad \quad + \left. {\FF( q_1^2, q_2^2) \ \FF( (q_1 + q_2)^2,
0) \over (q_1 + q_2)^2 - M_{\pi}^2} \ T_2(q_1,q_2;p) \right] \, . 
\label{a_pion_2}  
\eea
The pion-photon-photon transition form factor $\FF$ is defined by 
\be
i \int d^4 x e^{i q \cdot x} \langle \Omega | T \{ j_\mu(x) j_\nu(0)
\} | \pi^0(p) \rangle =\varepsilon_{\mu\nu\alpha\beta} \, q^\alpha
p^\beta \, \FF(q^2,(p-q)^2), \label{def_FF}
\ee
and $T_1$ and $T_2$ in Eq.~(\ref{a_pion_2}) are polynomials of up
to sixth order in the momenta $p,q_1,$ and $q_2$, with $p^2 = m^2$, see
Ref.~\cite{KN_pion} for details.  

Since no data on the doubly off-shell form factor $\FF(q_1^2,q_2^2)$
is available, one has to resort to theoretical descriptions of the
form factor. In order to proceed with the analytical evaluation of the
two-loop integrals in Eq.~(\ref{a_pion_2}), which is very difficult
without specifying $\FF(q_1^2,q_2^2)$, we considered a certain class
of form factors which includes the ones based on large-$N_C$ QCD that
we studied in Ref.~\cite{KN_VAP}.  For comparison, we have also used a
vector meson dominance (VMD) form factor (which leads to a result
close to the more sophisticated models used in Refs.~\cite{HKS,BPP}),
and a constant form factor, derived from the Wess-Zumino-Witten (WZW)
term, that describes the chiral anomaly in chiral perturbation
theory~\cite{WZW_ABJ}. In large-$N_C$ QCD, the pion-photon-photon form
factor has the form
\be
\FF(q_1^2,q_2^2)\big\vert_{N_C\to\infty}\,=\,
\sum_{ij}\frac{c_{ij}(q_1^2,q_2^2)}{(q_1^2-M_{V_i}^2)(q_2^2-M_{V_j}^2)}
\,,
\label{large_Nc_FF}
\ee
where the sum runs over an infinite set of narrow vector resonances.
The polynomials $c_{ij}(q_1^2,q_2^2)$ are arbitrary, however, there
are constraints at long and short distances. The normalization is
given by the WZW term, $\FF(0,0) = - N_C / (12 \pi^2 F_\pi)$, whereas
the OPE tells us that
\be
\lim_{\lambda\to \infty}\,\FF(\lambda^2 q^2, (p-\lambda q)^2)\,=\,
\frac{2}{3}\,\frac{F_\pi}{q^2}\,\bigg\{ \frac{1}{\lambda^2}\,+\,
\frac{1}{\lambda^3}\,\frac{q\cdot p}{q^2}\,+\,{\cal
O}\left({1 \over \lambda^4 }\right)\bigg\}\,. 
\label{OPE_FF}
\ee
In the following, we consider the form factors that are obtained by
truncation of the infinite sum (\ref{large_Nc_FF}) to one (lowest
meson dominance, LMD), respectively, two (LMD+V), vector resonances
per channel:
\bea
\FF^{\mbox{{\tiny LMD}}}(q_1^2,q_2^2) & = &  \!{F_\pi \over 3} {
q_1^2 + q_2^2 - c_V  \over (q_1^2 - M_V^2) (q_2^2 - M_V^2) }, 
\label{FF_LMD}\\
\FF^{\mbox{{\tiny LMD+V}}}(q_1^2,q_2^2) & = & \!{F_\pi \over 3} {
q_1^2 q_2^2 (q_1^2\!+\!q_2^2) + h_1 (q_1^2\!+\!q_2^2)^2 + h_2 q_1^2 q_2^2
+ h_5 (q_1^2\!+\!q_2^2) + h_7 \over (q_1^2 - M_{V_1}^2) (q_1^2 - M_{V_2}^2)
(q_2^2 - M_{V_1}^2) (q_2^2 - M_{V_2}^2)}, \nonumber \\
&& \label{FF_LMD+V} 
\eea
with $c_V \,= \, N_C M_V^4 / (4\pi^2 F_\pi^2) \, , h_7 \,= \, - N_C
M_{V_1}^4 M_{V_2}^4 / (4\pi^2 F_\pi^2)$.  Not all parameters in the
LMD+V form factor are fixed by the normalization and the leading term
in the OPE.  We have therefore determined the coefficients $h_1,h_2,$
and $h_5$ phenomenologically. According to
Refs.~\cite{Brodsky_Lepage,CLEO} the form factor $\FF(-Q^2,0)$ with
one photon on shell behaves like $1/Q^2$ for large spacelike momenta,
$Q^2=-q^2$. Whereas the LMD form factor does not have such a behavior,
it can be reproduced with the LMD+V ansatz, provided that $h_1 = 0$. A
fit of the LMD+V form factor to the CLEO data~\cite{CLEO}, with $h_1 =
0$, yields $h_5 = 6.93 \pm 0.26~\mbox{GeV}^4$, see
Ref.~\cite{KN_VAP}. Analyzing the experimental data~\cite{Alavi99} for
the decay $\pi^0 \to e^+ e^-$ in the context of Ref.~\cite{Pi_ll},
leads to $|h_2| \lapprox 20~\mbox{GeV}^2$.

The WZW form factor is given by $\FF^{\mbox{\tiny
WZW}}(q_1^2,q_2^2) = - N_C / (12 \pi^2 F_\pi)$ and the usual VMD form
factor reads 
\be
\FF^{\mbox{\tiny VMD}}(q_1^2,q_2^2) = - {N_C \over 12 \pi^2 F_\pi}
{M_V^2 \over (q_1^2 - M_V^2)} {M_V^2 \over (q_2^2 - M_V^2)} \, .
\label{FF_VMD}
\ee
Note that this form factor does not correctly reproduce the OPE
in Eq.~(\ref{OPE_FF}).   

\begin{figure}[h]
\epsfxsize=23pc % will enlarge or reduce the postscript figures based
%on the xsize 
\centerline{\epsfbox{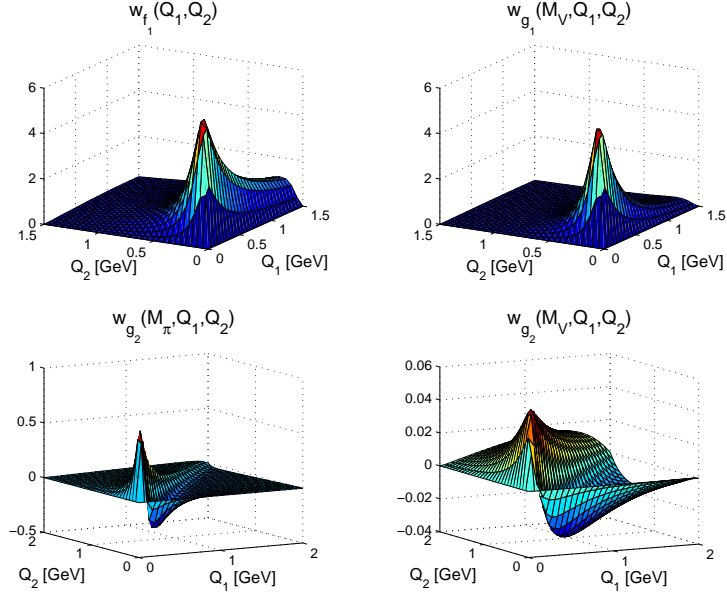}} % postscript image file name
\caption{The weight functions that appear in
Eqs.~\protect(\ref{api1}) and \protect(\ref{api2}). We have set $M_V =
M_\rho$.  
\label{fig:weightfunctions}} 
\end{figure}

All form factors above can be written in the following way 
\be \label{FF_f_g}
\FF(q_1^2, q_2^2) \,=\, \frac{F_\pi}{3}\,\bigg[ f(q_1^2)\,-\, 
\sum_{M_{V_i}} {1 \over q_2^2 - M_{V_i}^2}
g_{M_{V_i}}(q_1^2)\bigg] \, , 
\ee
where the explicit expressions for $f(q_1^2)$ and $g_{M_{V_i}}(q_1^2)$
for the different form factors can be found in~\cite{KN_pion}.  This
crucial observation allows one to cancel all dependences on $q_1
\cdot q_2$ in the numerators in $\FF(q_1^2,(q_1 + q_2)^2)$ in
Eq.~(\ref{a_pion_2}) and to perform all angular integrations in the
two-loop integrals analytically using the method of Gegenbauer
polynomials~\cite{Gegenbauer}. The pion-exchange contribution to $a_\mu$
can then be written as a two-dimensional integral representation as
follows, where the integration runs over the moduli of the Euclidean
momenta:
\bea
a_{\mu}^{\mbox{\tiny{LbyL;$\pi^0$}}} & = & 
\left( {\alpha \over \pi } \right)^3 
\left[ a_{\mu}^{\mbox{\tiny{LbyL;$\pi^0$}}(1)} + 
a_{\mu}^{\mbox{\tiny{LbyL;$\pi^0$}}(2)} \right] \, , \label{api_two_dim} \\
a_{\mu}^{\mbox{\tiny{LbyL;$\pi^0$}}(1)} 
& = & \int_0^\infty\!dQ_1 \int_0^\infty\!dQ_2  \Bigg[
w_{f_1}(Q_1,Q_2) \ f^{(1)}(Q_1^2,Q_2^2) \nonumber \\ 
&& \qquad \qquad \quad \   + \sum_{M_{V_i}}\!w_{g_1}(M_{V_i},Q_1,Q_2) \
g_{M_{V_i}}^{(1)}(Q_1^2, Q_2^2) \Bigg], \label{api1} \\
a_{\mu}^{\mbox{\tiny{LbyL;$\pi^0$}}(2)} 
& = & \int_0^\infty\!dQ_1 \int_0^\infty\!dQ_2\!\sum_{M=M_\pi,
M_{V_i}}\!w_{g_2}(M,Q_1,Q_2) \ g_{M}^{(2)}(Q_1^2, 
Q_2^2).  \label{api2}
\eea
The dependence on the form factors is contained in the functions
$f^{(1)},
g_{{\scriptscriptstyle M_{V_i}}}^{{\scriptscriptstyle (1)}},$ and
$g_{\scriptscriptstyle M}^{{\scriptscriptstyle (2)}}$ 
whereas the universal [for the class of form factors with the
decomposition~(\ref{FF_f_g})] weight functions $w_{f_1}, w_{g_1},
w_{g_2}$ are rational functions, square roots, and
logarithms~\cite{KN_pion}. We have plotted them in
Fig.~\ref{fig:weightfunctions}.
The functions $w_{f_1}$ and $w_{g_1}$ are positive and concentrated
around momenta of the order of $0.5~\mbox{GeV}$. 
Note, however, the tail in $w_{f_1}$ in the
$Q_1$ direction for $Q_2 \sim 0.2~\mbox{GeV}$, which produces for the
constant WZW form factor a divergence of the form $\sim {\cal C}
\ln^2 \Lambda$ for some UV-cutoff $\Lambda$. From the analytical
expression for $w_{f_1}$ we obtain~\cite{KN_pion} ${\cal C} = 3 [N_C m
/ (12 \pi F_\pi)]^2$. On the other hand, the function $w_{g_2}$ has
positive and negative contributions in the low-energy region, which
will lead to a strong cancellation in the corresponding integrals.
\begin{table}[h]
\caption{Results for the terms $a_{{\scriptscriptstyle
\mu}}^{\mbox{\tiny{LbyL;$\pi^0$}}(1)}$,  
$a_{{\scriptscriptstyle \mu}}^{\mbox{\tiny{LbyL;$\pi^0$}}(2)}$ and
$a_{{\scriptscriptstyle \mu}}^{\mbox{\tiny{LbyL;$\pi^0$}}}$ according to
Eq.~{\protect(\ref{api_two_dim})} for the different form factors. In
the WZW model we used a cutoff of $1~\mbox{GeV}$ in the first
contribution, whereas the second term is ultraviolet finite. 
In the LMD+V ansatz we have set $h_1 = 0~\mbox{GeV}^2$ and $h_5 =
6.93~\mbox{GeV}^4$.
} 
\begin{center}
\renewcommand{\arraystretch}{1.1}
\begin{tabular}{|l|r@{.}l|r@{.}l|r@{.}l|}
\hline
Form factor &
\multicolumn{2}{|c|}{{$a_{{\scriptscriptstyle
\mu}}^{\mbox{\tiny{LbyL;$\pi^0$}}(1)}$}}    
      & \multicolumn{2}{|c|}{{$a_{{\scriptscriptstyle
\mu}}^{\mbox{\tiny{LbyL;$\pi^0$}}(2)}$}}   
      & \multicolumn{2}{|c|}{{$a_{{\scriptscriptstyle
\mu}}^{\mbox{\tiny{LbyL;$\pi^0$}}} \times 10^{10}$}}   
\\ 
\hline  
WZW     & \hspace*{0.25cm} 0 & 095      & \hspace*{0.25cm}0 & 0020 &
\hspace*{0.65cm}  12 & 2 \\  
VMD     & 0 & 044       & 0 & 0013      & 5 & 6 \\ 
LMD     & 0 & 057       & 0 & 0014      & 7 & 3 \\
LMD+V ($h_2 = - 10~\mbox{GeV}^2$)
        & 0 & 049       & 0 & 0013      & 6 & 3 \\
LMD+V ($h_2 = 0~\mbox{GeV}^2$)
        & 0 & 045       & 0 & 0013      & 5 & 8 \\
LMD+V ($h_2 = 10~\mbox{GeV}^2$)
        & 0 & 041       & 0 & 0013      & 5 & 3 \\
\hline
\end{tabular}
\label{tab:api_models}  
\end{center}
\end{table}

In Table~\ref{tab:api_models} we present the numerical results for the
different form factors.  One observes that all form factors (apart
from the unrealistic constant WZW form factor, which only serves
illustrative purposes) lead to similar results, given mainly by
$a_{\mu}^{\mbox{\tiny{LbyL;$\pi^0$}}(1)}$. It is important
to correctly reproduce the slope of the form factor at the origin and
the data at intermediate energies. On the other hand, the asymptotic
behavior at large $Q_i$ seems not very relevant. All these
observations can easily be understood from the plots of the weight
functions. For a fixed value of $h_2$ in the LMD+V form factor, our
results are rather stable under the variation of the other
parameters. If we vary $M_{V_1}$, $M_{V_2}$, and $h_5$ by $\pm
20~\mbox{MeV}$, $\pm 25~\mbox{MeV}$, and $\pm 0.5~\mbox{GeV}^4$,
respectively, the result for $a_{\mu}^{\mbox{\tiny{LbyL;$\pi^0$}}}$
changes by $\pm 0.2 \times 10^{-10}$.  In contrast, if all other
parameters are kept fixed, our result depends for $|h_2| <
20~\mbox{GeV}^2$ almost linearly on $h_2$. In this range
$a_{\mu}^{\mbox{\tiny{LbyL;$\pi^0$}}}$ changes by $\pm 0.9 \times
10^{-10}$ from the value for $h_2 = 0$.

Thus, using the LMD+V form factor, our estimate reads 
\be 
a_{\mu}^{\mbox{\tiny{LbyL;$\pi^0$}}} = + 5.8~(1.0) \times 10^{-10}
\, , 
\ee
where the error includes the variation of the parameters and the
intrinsic model dependence. A similar short-distance analysis in the
framework of large-$N_C$ QCD and including quark mass corrections for
the form factors for the $\eta$ and $\eta^\prime$ was beyond the scope
of Ref.~\cite{KN_pion}. We therefore used VMD form factors fitted to
the relevant CLEO data~\cite{CLEO} to obtain our final estimate 
\be
a_{\mu}^{\mbox{\tiny{LbyL;PS}}} \equiv a_{\mu}^{\mbox{\tiny{LbyL;$\pi^0$}}}
+ a_{\mu}^{\mbox{\tiny{LbyL;$\eta$}}}\vert_{\mbox{\tiny VMD}} +
a_{\mu}^{\mbox{\tiny{LbyL;$\eta^\prime$}}}\vert_{\mbox{\tiny VMD}} = +
8.3~(1.2) \times 10^{-10} \, . 
\ee
We think that an error of 15~\% for the pseudoscalar pole contribution
is reasonable, since we impose many theoretical constraints from long
and short distances on the form factors. Furthermore, we use
experimental information whenever available, e.g.\ the CLEO data for
$\FF(Q^2,0)$ and the decay rate $\pi^0 \to e^+ e^-$. A better
measurement of the latter decay could considerably reduce the error
within the LMD+V ansatz, i.e.\ the bounds on the parameter $h_2$. Our
two-dimensional integral representation moreover shows that there are
no dangerous cancellations, at least in the main contribution
$a_{\mu}^{\mbox{\tiny{LbyL;$\pi^0$}}(1)}$.

% --------------------------------------------------------------------

\section{EFT approach to hadronic light-by-light scattering}
\label{sec:EFT}

In Ref.~\cite{a_mu_EFT} we discussed an EFT approach to
$a_{\mu}^{\mbox{\tiny{LbyL;had}}}$ based on an effective Lagrangian
that describes the physics of the Standard Model well below 1~GeV, see
also~\cite{EdeR_94}. It includes photons, light leptons, and the
pseudoscalar mesons and obeys chiral symmetry and $U(1)$ gauge
invariance. All pieces of $\lag_{\mbox{\small eff}}$ are available in
the literature~\cite{lag_eff} and generalizing the chiral power
counting by treating $e, m,$ and fermion bilinears 
as order $p$, one can write $\lag_{\mbox{\small eff}} = \lag^{(2)} +
\lag^{(4)} + \ldots,$ where the relevant terms are given by (for the
notation, see Ref.~\cite{lag_eff})
\bea
\lag^{(2)} &=& -\,\frac{1}{4}\,F_{\mu\nu}F^{\mu\nu} 
+ {\overline\psi}(i\not\!\!D - m)\psi
+ e^2C\langle QU^+QU \rangle  \nonumber \\
&&+ \frac{F_0^2}{4}\,\big(
\langle d^{\mu}U^+d_{\mu}U \rangle + 2B_0\langle {\cal M}^+U +
U^+{\cal M} \rangle \big) \,, \\
\lag^{(4)} &=& 
- {\alpha N_C \over 12 \pi F_0}
\varepsilon_{\mu\nu\alpha\beta} \, F^{\mu\nu} A^{\alpha}\partial^{\beta} 
\pi^0 + \cdots  \, ,  \label{L4WZW} \\
\lag^{(6)} &=& 
{\alpha^2 \over 4 \pi^2 F_0} \ \chi \ {\overline\psi} \gamma_\mu
\gamma_5 \psi \, \partial^\mu \pi^0 + \cdots . 
\label{L6chi}
\eea
 
The leading contribution to $a_{\mu}^{\mbox{\tiny{LbyL;had}}}$, of
order $\order(p^6)$, is given by a loop of charged pions with
point-like electromagnetic vertices. It is finite with the value $-
4.5 \times 10^{-10}$.~\cite{HKS} Since this contribution involves a
loop of hadrons, it is subleading in the large-$N_C$
expansion~\cite{EdeR_94}.

\begin{figure}[h]
\epsfxsize=18pc % will enlarge or reduce the postscript figures based
%on the xsize 
\centerline{\epsfbox{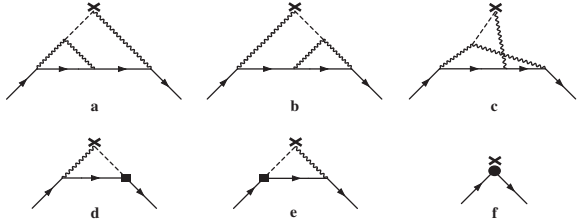}} % postscript image file name
\caption{The graphs contributing to
$a_{\mu}^{\mbox{\tiny{LbyL;$\pi^0$}}}$ at lowest order in the effective
field theory. 
\label{fig:EFT}}
\end{figure}

At leading order in $N_C$ and at order $p^8$ in $a_\mu$, we encounter
the divergent pion-pole contribution, diagrams (a) and (b) of
Fig.~\ref{fig:EFT}, involving two WZW vertices from Eq.~(\ref{L4WZW}).
Note that the diagram (c) is actually finite. The divergences of the
triangular subgraphs in the diagrams (a) and (b) are removed by
inserting the counterterm $\chi$ from Eq.~(\ref{L6chi}), see the
one-loop diagrams (d) and (e). Finally, there is an overall divergence
of the two-loop diagrams (a) and (b) that is removed by a local
counterterm denoted by $\kappa$, diagram (f). Note that at order $p^8$
there will also be additional contributions, if we replace the
point-like vertices in the charged pion-loop by vertices from
$\lag^{(4)}$, e.g.\ involving the low-energy constants $L_9$ and
$L_{10}$. These contributions will be subleading in $N_C$, although
they may be enhanced by logarithms.

\vspace*{-3mm} 
Since the EFT involves a local contribution $\kappa$ to $a_\mu$, we
will not be able to give a numerical prediction for
$a_{\mu}^{\mbox{\tiny{LbyL;had}}}$. Nevertheless, we can learn a few
things by considering the leading logarithms that are in addition
enhanced by a factor $N_C$ and which can be calculated using the
RG~\cite{Weinberg79}.  The expression of the renormalized contribution
to $a_{\mu}$ arising from the graphs of Fig.~\ref{fig:EFT} reads
\be \label{a_mu_EFT}
a_{\mu}^{\mbox{\tiny{LbyL;}}\pi^0}  =  
H(m / \mu ) + \chi(\mu) J(m / \mu) + \kappa(\mu)\,.  
\ee
The dependence on the subtraction scale $\mu$ in the two-loop function
reads $H(m/\mu) =\sum_{p=0,1,2} h_p \ln^p (m/\mu)$, and in the
one-loop function $J(m/\mu) = \sum_{q=0,1} j_q \ln^q(m/\mu)$.  The
physical quantity $a_{\mu}^{\mbox{\tiny{LbyL;$\pi^0$}}}$ satisfies
$\mu [d a_{\mu}^{\mbox{\tiny{LbyL;$\pi^0$}}} / d\mu]=0$. Since the
dimensionless coefficients $h_p$ and $j_q$ do not depend on $\mu$, but
are functions of the ratios $M_{\pi^0}/m$ and $F_{\pi}/m$, we have
${\cal D} h_p=0$ and ${\cal D} j_q=0$, where ${\cal D} \equiv m
\partial / \partial m + M_{\pi^0} \partial /
\partial M_{\pi^0} + F_{\pi} \partial / \partial F_{\pi}$. Finally,
the conditions ${\cal D}\chi(\mu) = 0$ and ${\cal D} \kappa(\mu) = 0$
hold. We thus obtain the two RG equations 
\be
h_2 = {1 \over 2} \gamma_\chi j_1 \, , \qquad
\mu\,\frac{d\kappa(\mu)}{d\mu} =  - \gamma_\chi j_0 + \chi(\mu) j_1
+ h_1 \, , \label{RG_eq} 
\ee
with $\gamma_{\chi} \equiv \mu [d \chi(\mu) / d\mu] = N_C$, see
Ref.~\cite{Pi_ll}. The coefficient $j_1$ of the logarithm in the sum
of the one-loop graphs (d) and (e) can easily be worked
out~\cite{a_mu_EFT}. In this way we obtain from Eq.~(\ref{RG_eq}) 
\be
h_2\equiv \left( {\alpha \over \pi} \right)^3 \, {\cal C} \, , \quad 
{\cal C} =
+3\,\left(\frac{N_C}{12\pi}\right)^2\,\left(\frac{m}{F_{\pi}}\right)^2 
,  \label{Clog2}
\ee
in agreement with the result given in Section~\ref{sec:pionpole}. 

% --------------------------------------------------------------------

\section{Further comments and conclusions}
\label{sec:comments_conclusions}

The EFT and large-$N_C$ analysis tells us that we can write 
\bea
a_{\mu}^{\mbox{\tiny{LbyL;had}}} & = &
\left( {\alpha \over \pi} \right)^3  \Bigg\{
f\left({m_{\pi^\pm} \over m}, {m_{K^\pm} \over m}\right)
\nonumber \\
&&\qquad\quad
+ N_C \left( {m^2 \over 16 \pi^2
F_\pi^2} {N_C \over 3} \right)
\left[ \ln^2 {\mu_0 \over m} + c_1 \ln {\mu_0 \over m} + c_0
\right]  \nonumber \\ 
&&\qquad\quad+ \order \left( {m^2 \over \mu_0^2} \times
\mbox{log's}\right) + \order \left( {m^4 \over \mu_0^4} \times
N_C \times \mbox{log's}\right)  \Bigg\} 
\, , \label{a_mu_EFT_N_C}
\eea
where~\cite{HKS} $f(m_{\pi^\pm}/m, m_{K^\pm} / m) = - 0.038$ 
represents the charged pion and kaon-loop that is formally of
order one in the chiral and $N_C$ counting. 
We have denoted by $\mu_0$ some hadronic scale, e.g.\ $M_\rho$. 
The coefficient of the log-square term in the second line is universal
and of order $N_C$, since $F_\pi = \order(\sqrt{N_C})$. The remaining
parts of the coefficient $c_1$ have recently been calculated in
Ref.~\cite{Ramsey-Musolf_Wise} in the $\overline{\mbox{MS}}$-scheme:
$c_1 = - [2 \chi(\mu_0) / 3 - 0.904]$. The low-energy
constant $\chi$ can be determined from the decay $\pi^0
\to e^+ e^-$. However, due to the big experimental error, this leads
to a large uncertainty in $\chi$ and thus in
$a_{\mu}^{\mbox{\tiny{LbyL;had}}}$. Assuming lepton universality, one
can obtain $\chi$ also from the decay $\eta \to
\mu^+ \mu^-$, with better precision $\chi(M_\rho) =
1.75^{+1.25}_{-1.00}$.~\cite{Ametller} One observes that although
the logarithms are sizeable, $\ln(M_\rho / m) = 1.98$, and give the
ballpark of the final result, there is some cancellation between the
log-square and the log-term in $a_{\mu}^{\mbox{\tiny{LbyL;had}}}$.
Note that the form factors based on large-$N_C$ QCD that we discussed
in Section~\ref{sec:pionpole} give values for the constant $c_0$ in
Eq.~(\ref{a_mu_EFT_N_C}) that are of order one, i.e.\ of natural
size. Furthermore, in addition to $\pi^0 \to e^+ e^-$, we have also
taken other experimental and theoretical constraints on $\FF$ into
account. 

It has been argued that $a_{\mu}^{\mbox{\tiny{LbyL;had}}}$ has to be
positive, because of quark-hadron duality, since also the
(constituent) quark-loop leads to a positive result. We do not agree
with this argument for the following
reason. Equation~(\ref{a_mu_EFT_N_C}) tells us that at leading order
in $N_C$ any effective theory or model of QCD has to show the behavior
$a_\mu^{\mbox{\tiny{LbyL;had}}} \sim (\alpha/\pi)^3 N_C [N_C m^2 / (48
\pi^2 F_\pi^2)] \ln^2\Lambda$, with a universal coefficient, 
if one sends the cutoff $\Lambda$ to infinity. From the analytical
result given in Ref.~\cite{Laporta_Remiddi} for the quark-loop one
obtains the behavior $a_\mu^{\mbox{\tiny LbyL;CQM}} \sim
(\alpha/\pi)^3 N_C (m^2 / M_Q^2) + \ldots$, for $M_Q \gg m$, if we
interpret the constituent quark mass $M_Q$ as a hadronic cutoff. Note
that the quark-loop is also leading in large $N_C$.~\cite{EdeR_94}
Even though one may argue that $N_C / 16 \pi^2 F_\pi^2$ in ${\cal C}$
can be replaced by $1/M_Q^2$, the log-square term is not correctly
reproduced with this model. Therefore, the constituent quark model
cannot serve as a reliable description for the dominant contribution
to $a_\mu^{\mbox{\tiny{LbyL;had}}}$. On the other hand, the
coefficient ${\cal C}$ could, a priori, have {\it any sign}, since it
is determined purely within the EFT, where it is related to the decay
amplitude $\pi^0 \to e^+ e^-$.

In conclusion, the pseudoscalar exchange contribution
$a_{\mu}^{\mbox{\tiny{LbyL;PS}}}$ seems to be under control at the
15~\% level. In addition, the EFT and large-$N_C$ analysis shows a
systematic approach to $a_{\mu}^{\mbox{\tiny{LbyL;had}}}$ and yields
the leading and next-to-leading logarithmic terms, enhanced by a
factor $N_C$. On the other hand, these terms tend to cancel each other
to some extent. Furthermore there remains the issue of the other
contributions to $a_{\mu}^{\mbox{\tiny{LbyL;had}}}$. For instance, the
charged pion-loop is the leading term in the low-energy expansion,
however, after dressing the couplings to the photons with form
factors, there is a huge suppression, by a factor of 3 to 10, although
the model-dependent results after the dressing differ by about $1
\times 10^{-10}$.~\cite{HKS,BPP} Similar statements apply to the
dressed quark-loop, that appears in these models.  Taking all these
uncertainties into account by adding the model-dependent errors
linearly, my conservative estimate for the full hadronic
light-by-light scattering contribution to $a_\mu$ is as follows:
\be
a_{\mu}^{\mbox{\tiny{LbyL;had}}} = +~8~(4) \times 10^{-10} \, . 
\ee

% --------------------------------------------------------------------

\section*{Acknowledgments}
\vspace*{-1mm} 
I would like to thank Marc Knecht, Michel Perrottet, and Eduardo de
Rafael for the pleasant collaboration on the topics presented here,
Richard Lebed for the invitation to the conference, and the Institute
for Nuclear Theory in Seattle and the Schweizerischer Nationalfonds
for financial support.

\end{document}